\documentclass[PRL,showpacs,amsmath,amssymb,superscriptaddress,twocolumn]{revtex4}
\usepackage[T1]{fontenc}
\usepackage[latin1]{inputenc}
\usepackage{amsmath}
\usepackage{graphicx}
\usepackage{epsfig}
\usepackage{bm}
\usepackage{dsfont}
\usepackage{color}

\newcommand{\nn}{\nonumber}

\newcommand{\bea}{\begin{eqnarray}}
\newcommand{\ea}{\end{eqnarray}}
\newcommand{\beq}{\begin{equation}}
\newcommand{\eq}{\end{equation}}
\newcommand{\bc}{\begin{center}}
\newcommand{\ec}{\end{center}}
\newcommand{\dg}{\dagger}
\newcommand{\la}{\langle}
\newcommand{\ra}{\rangle}
\newcommand{\ov}{\overline}

\begin{document}

\title{Inflationary quasiparticle creation and thermalization dynamics in coupled Bose-Einstein condensates}

\author{Anna Posazhennikova}
\email[Email: ]{anna.posazhennikova@rhul.ac.uk}

\affiliation{Department of Physics, Royal Holloway, University of London,
  Egham, Surrey TW20 0EX, United Kingdom}

\author{Mauricio Trujillo-Martinez}
\affiliation{Physikalisches Institut and Bethe Center for Theoretical Physics,
  Universit\"at Bonn, Nussallee, 12, D-53115 Bonn, 
Germany}

\author{Johann Kroha}
\email[Email: ]{kroha@th.physik.uni-bonn.de}

\affiliation{Physikalisches Institut and Bethe Center for Theoretical Physics,
  Universit\"at Bonn, Nussallee, 12, D-53115 Bonn, 
Germany} 

\affiliation{Center for Correlated Matter, Zhejiang University, 
Hangzhou, Zhejiang 310058, China} 

\received{13 April 2016; accepted 10 May 2016}

\begin{abstract}
A Bose gas in a double-well potential, exhibiting a true Bose-Einstein 
condensate (BEC) amplitude and initially performing Josephson oscillations,
is a prototype of an isolated, non-equilibrium many-body system.
We investigate the quasiparticle (QP) creation and thermalization dynamics of
this system by solving the time-dependent Keldysh-Bogoliubov equations.
We find avalanche-like QP creation due to a parametric resonance between 
BEC and QP oscillations, followed by  slow, exponential relaxation to a 
thermal state at an elevated temperature, controlled by the initial excitation 
energy of the oscillating BEC above its ground state. The crossover 
between the two regimes occurs because of an effective decoupling of the QP and 
BEC oscillations. This dynamics is analogous to elementary particle 
creation in models of the early universe. The thermalization in our set-up occurs because 
the BEC acts as a grand canonical reservoir for the quasiparticle system.
\end{abstract}


\maketitle

Common knowledge tells that many-body systems come to thermodynamic 
equilibrium by coupling to a heat reservoir. However, how can an 
{\it isolated} quantum many-body system eventually come to rest from a given initial 
non-equilibrium  state, and is the final state a thermal one?
This is a long-standing problem, which has recently received intense interest 
\cite{Polkovnikov2011,Eisert2015}, inspired by the
high degree of isolation and control possible in ultracold quantum gases
\cite{Trotzky2012,Gring2012}. While the unitary time evolution of 
an isolated quantum system rigorously prohibits the maximization of the 
total entropy and, thus, effective thermalization is 
generically observed \cite{Trotzky2012,Gring2012}.
 
Several mechanisms have been put forward in order to 
resolve this contradiction, most notably the eigenstate  
thermalization hypothesis (ETH) \cite{Deutsch1991,Srednicki1994}. 
It conjectures that for a sufficiently complex quantum system the thermal 
average of an observable at a given average energy is practically 
indistinguishable from its expectation value in an 
eigenstate of the system with that energy. The ETH has been verified 
numerically for generic, non-integrable systems 
and typical observables \cite{Rigol2008,Rigol2012}, and was found to fail for
integrable systems \cite{Rigol2009,Rigol2012} with some 
exceptions \cite{Alba2015}. Another mechanism may be termed 
subsystem thermalization hypothesis (STH). It relies on 
the fact that, even though total entropy maximization is not possible, 
subsystems may thermalize by exchanging energy and/or particles 
with other parts of the system, so that averages of local quantities 
may be thermal. This mechanism has been successfully invoked 
\cite{Essler2014} even for
integrable and nearly integrable systems exhibiting pre-thermalization dynamics 
\cite{Kollath2007,Kehrein2008,Kollar2011,Essler2014}. 
The STH is also at the heart of hydrodynamic behavior, 
where physical quantities first relax to local averages and then evolve 
slowly, under the rule of local conservation laws. 
However, a unified understanding of thermalization has not been reached,
and the thermalization mechanism seems to depend strongly on the type of 
system 
\cite{Rigol2012,Rigol2009,Alba2015,Essler2014,Kollath2007,Kehrein2008,Kollar2011,
Schollwoeck2008,Yukalov2011,Eisert2012}.

In the present work we investigate the thermalization dynamics of an
interacting Bose gas trapped in a double-well potential which supports 
a true Bose-Einstein condensate (BEC), initially performing non-equilibrium 
Josephson oscillations between the two wells 
\cite{Smerzi1997,Albiez2005,Levy2007,Thywissen2011}. 
This is a prototype of a non-integrable system with a natural subsystem
structure, namely the BEC and the system of incoherent excitations
of Bogoliubov quasiparticles (QPs). The existence of a true BEC phase 
precludes the system to be one-dimensional. Therefore, and because of the 
large particle number considered, numerically exact methods, like the 
time-dependent density-matrix renormalization group (t-DMRG) 
\cite{Kollath2007,Schollwoeck2008},
are not applicable here. Moreover, the slow thermalization dynamics found
below requires evolving the system to large times, difficult to reach by 
these methods. Instead, the non-equilibrium Keldysh-Bogoliubov 
formalism in the grand-canonical ensemble is appropriate here, since 
the BEC acts as a particle reservoir for the QP system (and vice versa).

We find rich dynamics, governed by three different time scales. 
After an initial period of undamped Josephson oscillations
\cite{Mauro2009,Mauro2015}, 
QPs are created in an avalanche manner (qp creation time, $\tau_c$) due to a 
dynamically generated parametric resonance between the Josephson frequency 
and the QP excitation energies. This leads to a fast depletion as well as 
damping of the BEC amplitude \cite{Yukalov2008}. 
When the final number of QP excitations, 
allowed by total energy conservation, is reached, however, 
the QP system effectively decouples from the BEC oscillations 
(freeze-out time of the BEC, $\tau_f$), and the total QP number becomes nearly
conserved. Under this approximate conservation law the system enters 
into a quasi-hydrodynamic regime which is characterized by slow, 
exponential relaxation of the QP system into a thermalized state 
(thermalization time, $\tau_{th}$). We prove this behavior by a 
detailed spectral analysis of the oscillatory behavior in the different 
time regimes.        

{\it Model and formalism.}~-- 
The Bose gas is described by the Hamiltonian 
\bea
H=\int d {\bf r} \hat \Psi^{\dg}({\bf r},t)\left(-\frac{\nabla^2}{2m}+V_{ext}({\bf
    r},t)\right)\hat \Psi({\bf r},t) \\ \nonumber
+\frac{g}{2}\int d {\bf r} \hat \Psi^{\dg}({\bf r},t)\hat \Psi^{\dg}({\bf
  r},t)\hat \Psi({\bf r},t)\hat \Psi({\bf r},t),
\label{gen_ham}
\ea
where $\hat \Psi({\bf r},t)$ is a bosonic field operator, and $g=4\pi a_s/m$
is a contact interaction constant, with $a_s$ the s-wave scattering length. 
$V_{ext}$ is the external double-well trap potential.
This system is known to exhibit 
Josephson oscillations \cite{Smerzi1997,Albiez2005, Levy2007}.
In our approach the condensate is described within a semiclassical two-mode 
approximation \cite{Smerzi1997}, while the QP dynamics are 
described quantum-mechanically \cite{Mauro2009,Mauro2015}.
We now represent 
$\hat \Psi({\bf r},t)$  in terms of the 
complete basis  
$\mathds{B}=\{\varphi_-,\varphi_+,\varphi_1,\varphi_2, \dots \varphi_M  \}$  
of the exact single-particle eigenstates of $V_{ext}({\bf r})$ after 
the coupling between the wells is turned on at $t=0$ by suddenly lowering the barrier between the wells \cite{Mauro2015}.  Hence for 
times $t>0$
\beq
\hat \Psi({\bf r},t)=\phi_1({\bf r}) a_1(t)+\phi_2({\bf
  r}) a_2(t)+\sum_{n=1}^{M}\varphi_n({\bf r}) \hat b_n(t)\ ,
\label{field_operator2}
\eq
The first two terms in Eq. (\ref{field_operator2}) constitute the usual 
two-mode approximation \cite{Smerzi1997}, 
i.e., $\phi_1$ and $\phi_2$ are symmetric and antisymmetric superpositions 
of $\varphi_-$ and $\varphi_+$, the ground state and the first excited state
of $V_{ext}({\bf r})$. Hence, the wave function of $\phi_1$ ($\phi_2$) is
localized in the left (right) potential well, and
$a_{\alpha}(t)=\sqrt{N_{\alpha}(t)}\exp\left({i\theta_{\alpha}(t)}\right)$, 
$\alpha=1,2$, are the corresponding BEC amplitudes.  
This Bogoliubov substitution neglects phase fluctuations 
in the ground states of each of the potential wells which is justified 
for sufficiently large BEC particle numbers, 
$N_{\alpha}(t)\gg 1$, e.g., for the experiments \cite{Albiez2005}. 
For the excited states, $\varphi_n$, $n=1,\ 2,\dots,\ M$, the full 
quantum dynamics are taken into account by the bosonic creation and 
destruction operators $\hat b_n^{\dagger}$, $\hat b_n^{\phantom{\dagger}}$. 

 For $t>0$ the 
Hamiltonian of our system is  $H=H_{coh}+H_{J}+H_{coll}$.
$H_{coh}$ includes all coherent, local contributions, i.e., all terms
which are bi-linear in the $\hat b_n$-operators and local in 
the well index $\alpha = 1,\ 2$, 
\bea
H_{coh}=\varepsilon_0\sum_{\alpha=1}^2a_{\alpha}^*a_\alpha+\frac{U}{2}\sum_{\alpha=1}^2a_{\alpha}^*a_{\alpha}^*a_\alpha a_\alpha+\sum_{n=1}^M \varepsilon_n\hat b_n^{\dagger} \hat b_n \nn \\
+K\sum_{\alpha=1}^2\sum_{n,m=1}^M \left[a_{\alpha}^*a_{\alpha}\hat b_n^{\dg}\hat b_m+\frac{1}{4}(a_{\alpha}^*a_{\alpha}^*\hat b_n \hat b_m+h.c.)\right],
\ea
where $U$ and $K$ are positive interaction constants, and $\varepsilon_n$ are 
the energies of the $M$ equidistant levels of the double well,
separated by the trap frequency, $\varepsilon_n=n\Delta$. For simplicity we 
neglect here and in the following a possible level-dependence of the 
coupling constants.

$H_J$ encompasses the Josephson terms, which are still coherent but non-local
in the well index, 
\bea
H_{J}&=&-J(a_1^*a_2+a_2^*a_1)+J'\sum_{n,m=1}^M \left[ (a_1^*a_2+a_2^*a_1)\hat b_n^{\dg}\hat b_m \right. \nn \\
&+&\left.\frac{1}{2}(a_1^*a_2^*\hat b_n\hat b_m +h.c.) \right] \ .
\ea
The terms proportional to $J'$ constitute QP-assisted tunneling between the wells.

Finally, the non-linear collisional terms $H_{coll}$ account for 
QP interactions, 
\bea
H_{coll}=\frac{U'}{2}\sum_{n,m=1} \sum_{l,s=1} \hat b_m^{\dagger}  \hat
b_n^{\dagger} \hat b_l \hat b_s \hspace*{3.3cm}  
\ea
\vspace*{-0.4cm}
\bea 
+R\hspace*{-0.06cm}\left[\sum_{\alpha=1}^2\sum_{n,m,s=1} a_{\alpha}^*\hat
  b_n^{\dagger} \hat b_m\hat b_s + \hspace*{-0.06cm} 
  \sum_{\alpha,\beta,\gamma+1}^2\sum_{n=1} a^*_{\alpha}a^*_{\beta}a_{\gamma}\hat b_n +h.c.\right] \nn
\label{H_collisions}
\ea

The time evolution of this system is described in terms of the 
condensate population imbalance, $z(t)=\frac{N_1(t)-N_2(t)}{N_1(t)+N_2(t)}$, 
the phase difference between the BECs, $\theta(t)$
and the QP occupation numbers $n_1(t),\ n_2(t),\ \dots, n_M(t)$. They 
can be calculated from the classical ${\bf C}$ and the quantum ${\bf G}$ parts 
of the two-time Green's functions following standard field-theoretical 
techniques \cite{Rammerbook,Griffinbook},
\bea
{\bf C}_{\alpha\beta}(t_1,t_2)\hspace*{-0.12cm}&=&\hspace*{-0.12cm}
-i\left( 
\begin{matrix}
a_\alpha(t_1)a_\beta^*(t_2) & a_\alpha(t_1) a_\beta(t_2) \\
a_\alpha^*(t_1)a_\beta^*(t_2)  & a_\alpha^*(t_1) a_\beta(t_2)
\end{matrix}
\right) 
\label{SM:cond}\\
{\bf G}_{nm}(t_1,t_2)\hspace*{-0.12cm}&=&\hspace*{-0.12cm}
-i\left( 
\begin{matrix}
\la T_C\hat b_n(t_1)\hat b_m^\dg (t_2) \ra & \la T_C\hat b_n(t_1)\hat b_m (t_2) \ra \\ \nn
\la T_C\hat b_n^\dg(t_1)\hat b_m^\dg (t_2) \ra  & \la T_C\hat b_n^\dg(t_1)\hat b_m (t_2) \ra
\end{matrix}
\right)  
\label{SM:GF_qp}
\ea
where $\hat T_C$ denotes time-ordering along the Keldysh contour. 
The Dyson equations for these functions read,
\bea
\int_C d\ov t \left[{\bf G }_0 ^{-1} (t_1,\ov t)\right. 
&-& \left. {\bf S}^{HF} (t_1,\ov t)
  \right] {\bf C} (\ov t, t_2 )   \nn \\
 &=& \int_C d\ov t \,{\bf S }(t_1,\ov t)  {\bf C} (\ov t, t_2 )
\label{dyson-cond-keldysh} 
\ea
\bea
\int_C d \ov t \left[{\bf G }_0 ^{-1} (t_1,\ov t) \right. &-& \left. 
{\bf \Sigma}^{HF} (t_1,\ov t) \right] {\bf G} (\ov t, t_2 ) \nn \\
 = \mathds{1} \delta(t_1-t_2)  &+& \int_C d \ov t\, {\bf \Sigma } (t_1,\ov t)
 {\bf G} (\ov t, t_2 ) \label{dyson-noncond-keldysh} \ .
\ea
In Eqs. (\ref{dyson-cond-keldysh}), (\ref{dyson-noncond-keldysh}), 
the first-order (Hartree-Fock) self-energies ${\bf S}^{HF},{\bf \Sigma}^{HF}$ 
describe time-dependent level renormalizations, while the second-order 
collisional self-energy contributions ${\bf S},{\bf \Sigma}$
induce damping of the QP and BEC oscillations.  
The Dyson equations are expressed 
in terms of the spectral function 
${{\bf A}_{nm}}(t_1,t_2)=i({\bf G}_{nm}^>(t_1,t_2)-{\bf G}_{nm}^<(t_1,t_2))$ 
and the statistical function 
${{\bf F}_{nm}}(t_1,t_2)=({\bf G}_{nm}^>(t_1,t_2)+{\bf G}_{nm}^<(t_1,t_2))/2=
{\bf G}^K_{nm}(t_1,t_2)/2$ and the corresponding self-energies 
(see Supplementary
Material). We solve the resulting integro-differential equations numerically for
total number of particles $N_{\text{tot}}$, level spacing $\Delta$,  
interactions $U,\,U',\,K,\,J',\,R$ and  
initial conditions $z(0), \theta(0)$, with all particles being initially in
the BEC, $N_1(0)+N_2(0)=N_{\text{tot}}$. 
All energies are expressed in units of $J$: $u=UN_{\text{tot}}/J$,
$u'=U'N_{\text{tot}}/J$, $k=KN_{\text{tot}}/J$, $j'=J'N_{\text{tot}}/J$, 
$r=RN_{\text{tot}}/J$. In the numerical evaluations we limit the 
number of levels which can be occupied by the QPs to $M=5$.

\begin{figure}[!bt]
\begin{center}
\includegraphics[width=0.45\textwidth]{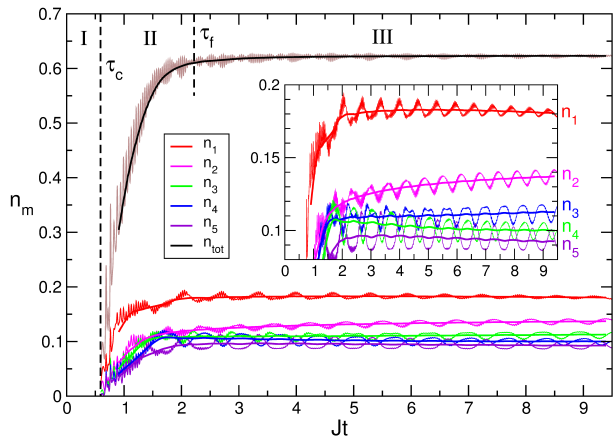}\vspace*{-0.8em} 
\end{center}
\caption{Dynamics of incoherent excitations for $z(0)=0.6$, $\theta(0)=0$, and
  $\Delta=9$, $u=u'=5$, $j'=40$, r=300, $N_{\text{tot}}=5\cdot 10^5$. $n_m$ is the
  occupation number of the $m-$th level, $n_{\text{tot}}$ is the sum of all
  $M=5$ levels. All occupation numbers shown are normalized by the total 
  particle number $N_{\text{tot}}$. The three different dynamical regimes,
  separated by the characteristic times $\tau_c$ and $\tau_f$, are marked,
  as explained in the text. The inset shows an enlargement of the long-time
  behavior.}
\label{qp_dynamics}
\end{figure} 

\begin{figure}[!bt]
\begin{center}
\includegraphics[width=0.45\textwidth]{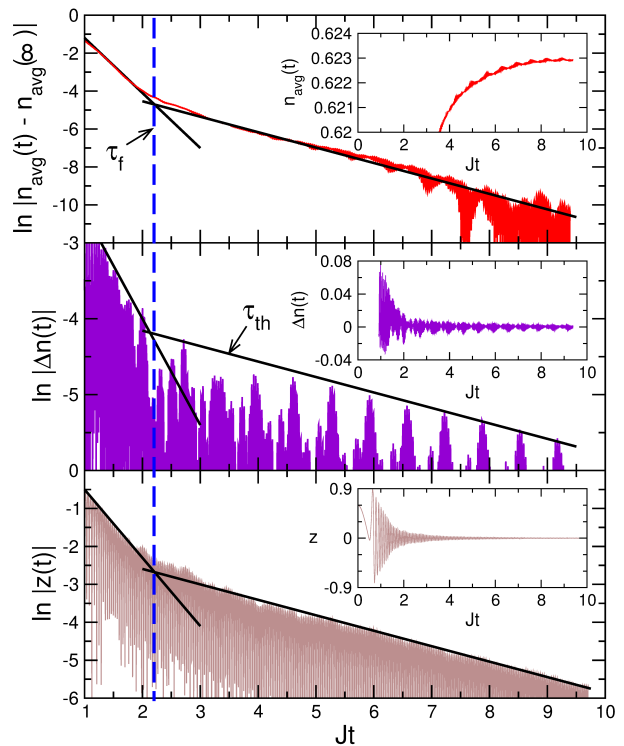}\vspace*{-0.8em} 
\end{center}
\caption{Logarithmic plots of the relaxation behavior of the QP system 
(upper two panels) and of the BEC population imballance (lower panel),
see text. The dashed vertical line marks the freeze-out time $\tau_f$ where the 
BEC system and the system of incoherent excitations effectively decouple.
The thin, black lines are guides to the eye. 
The insets show the respective linear plots, for illustration.}
\label{log}
\end{figure} 
 
{\it BEC and QP dynamics.}~-- 
Fig. \ref{qp_dynamics} 
shows the dynamics of incoherent excitations for a typical parameter set, 
given in the figure caption. The time-dependent occupation numbers
of all $M=5$ levels, $n_1,\,n_2,\,\dots ,\,n_M$, and their total,   
\beq 
n_{\text{tot}}(t)=\sum_{m=1}^M n_m(t)=-\sum_{m=1}^M \left[{\rm Im} \text{F}^G_{mm}(t,t) -\frac{1}{2}\right]\ ,
\eq
are shown.  
Here $\text{F}^G_{mm}$ is the regular (upper diagonal) component of the
equal-time statistical Green's function ${{\bf F}_{mm}}$ in Bogoliubov space 
(see Supplementary Material). 
From Fig. \ref{qp_dynamics} one can readily identify three different 
dynamical regimes, (I) an early regime of undamped Josephson oscillations 
without QPs for $t<\tau_c$ \cite{Mauro2009,Mauro2015}, (II) a fast growth regime of the QP population, 
and (III) a regime of slow relaxation to a stationary state for long times. 
In the regimes (II) and (III) the $n_m(t)$ and 
$n_{\text{tot}}(t)$ oscillate around their respective running mean values, 
$n_{m,\,\text{avg}}(t)$ and $n_{\text{avg}}(t)$ (averaged over one oscillation period; 
smooth lines on top of the oscillating ones).   
To analyze the functional dependence of this time evolution, we show in 
Fig.~\ref{log} logarithmic plots of the deviation of the total running mean 
$n_{\text{avg}}(t)$ from its final value $n_{\text{avg}}(\infty)$ (upper
panel) and the momentary oscillation amplitude 
$\Delta n(t)=n_{\text{tot}}(t)-n_{\text{avg}}(t)$ 
(middle panel) along with the BEC population imballance $z(t)$ 
(lower panel). All three quantities show a steep crossover from the 
fast growth regime (II) to the slow relaxation regime (III) at a 
freeze-out time scale $\tau_f$, 
with exponential relaxation for $t>\tau_f$.

\begin{figure}[!bt]
\begin{center}
\includegraphics[width=0.45\textwidth]{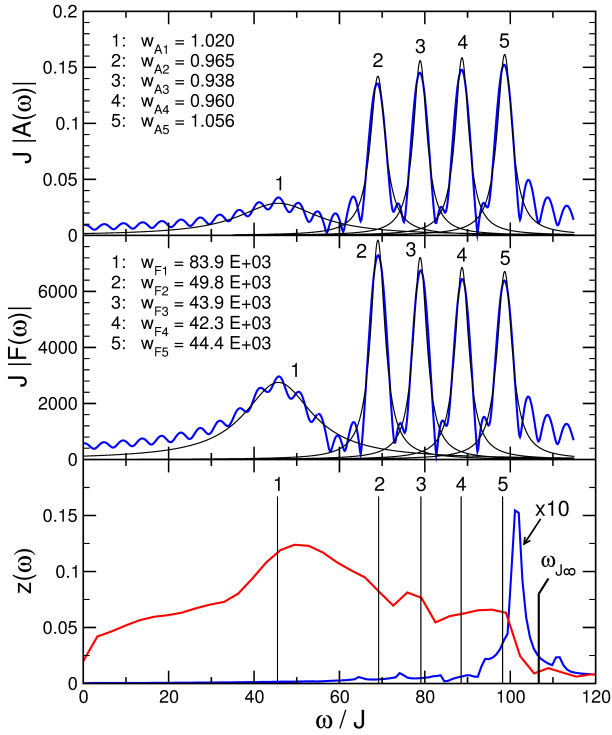}\vspace*{-0.8em} 
\end{center}
\caption{Absolute values of spectral (upper panel) and statistical (middle
  panel) functions, Fourier-transformed with respect to $\tau=(t_1-t_2)$ for a
  fixed value of $t=(t_1+t_2)/2=9.01/J$. The thin, black lines represent 
  Lorentzian fits. The weights $\text{w}$ of each of the $M=5$ 
  Lorentzians are shown in the insets. In the lower panel the power 
  spectrum $z(\omega)$ of the BEC population imballance is shown for 
  $\tau_c\lesssim t \lesssim \tau_f$ (red line) and for $t>\tau_f$ (blue
  line). The vertical lines indicate renormalized QP energies. 
  $\omega_{J\infty}$ is the Josephson frequency estimated for the 
  decoupled, quasi-hydrodynamic regime (see text). }
\label{spectra}
\end{figure}

What are the mechanisms for fast growth (II) and slow relaxation (III),
and how is the steep crossover time $\tau_f$ determined? 
A spectral analysis provides detailed insight into these problems.
We introduce the usual Wigner ``center-of motion'' (CoM) time 
$t=(t_1+t_2)/2$ and difference time $\tau =(t_1 - t_2)$ and Fourier-transform 
the two-time Green's functions 
$\text{A}^G(t_1,t_2)=\sum_n\text{A}_{nn}^G(t_1,t_2)$ 
and $\text{F}^G(t_1,t_2)=\sum_n\text{F}_{nn}^G(t_1,t_2)$ with respect to
$\tau$. 
In Fig. \ref{spectra} (upper and middle panels) we plot the 
frequency-dependent absolute values of
$\text{A}^G(\omega,t)\equiv \text{A}(\omega)$ and
$\text{F}^G(\omega,t)\equiv \text{F}(\omega)$ in the long-time regime,
$t=9.01 /J >\tau_f$.  
As expected, the spectra exhibit 
$M=5$ approximately Lorentzian peaks corresponding to the five 
renormalized QP levels. They mark the Rabi 
oscillation frequencies of the non-equilibrium QP system.
Note that, at any instant $t$ of the time evolution, the 
maximum time interval available for $\tau$ is necessarily finite, 
$-2t <\tau < 2t$ (see Supplementary Material). This limits the frequency 
resolution of the Fourier Transform to $2\pi/4t$ and results in the wiggly modulations of the 
Lorentzian peaks. For $\tau_c<t<\tau_f$ the spectra look similar, however with
reduced $\omega-$resolution (not shown).  
Fig. \ref{spectra} (lower panel) displays the power spectra of the 
BEC population imballance $z(t)$, Fourier transformed 
with respect to $t$ for $\tau_c<t<\tau_f$ (red curve, regime (II)) 
and for $t>\tau_f$ (blue curve, enlarged by a factor 10, regime (III)), 
respectively.

\begin{figure}[!bt]
\begin{center}
\includegraphics[width=0.45\textwidth]{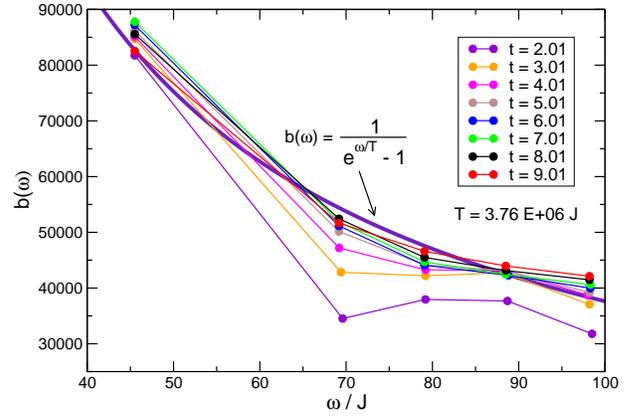}\vspace*{-0.8em} 
\end{center}
\caption{Distribution function $b(\tilde\varepsilon_n,t)$ for different
  CoM times $t$. The thick purple line is a single-parameter
  fit of a thermal distribution to the calculated 
  $b(\tilde\varepsilon_n,t)$  for the largest time $t=9.01$, 
  with temperatue $T$ as fit parameter. The fitted value is 
  $T=3.76\cdot 10^{6}\ J$.}
\label{distribution}
\end{figure}

{\it Inflationary QP creation.}~--
In the fast growth regime (II), the BEC oscillation spectrum $z(\omega)$ 
overlaps strongly with the QP spectrum $|\text{A}(\omega)|$ and has even maxima 
at the renormalized Rabi frequencies. This signals a dynamically 
generated, parametric resonance, and the strong, resonant QP-BEC coupling leads 
to the inflationary QP generation observed in Fig.~\ref{qp_dynamics}.

{\it Approach to stationary state.}~--
In the long-time regime (III), the behaviour is strikingly different:
the spectrum of the BEC oscillations exhibits a single,
sharp peak of substantially reduced weight which has almost no 
overlap with the QP spectrum. In fact, the BEC oscillation 
frequency is close to the Josephson frequency $\omega_{J\infty}$ 
of a semiclassical, interacting BEC, i.e. {\it without} locking 
to the QP oscillations. $\omega_{j_\infty}$ may be estimated as \cite{Smerzi1997}, 
$\omega_{J\infty}\approx 2J_{eff}\sqrt{1+uJ/(2J_{eff})}$, 
where $J_{eff}=J+n_{\text{tot}}(t\to\infty)J'$ is the QP-enhanced Josephson 
coupling. It is marked in Fig.~\ref{spectra} by the thick, vertical line.
Therefore, in the long-time regime the BEC system performs nearly free, weakly 
damped (due to the sharpness of the spectral peak) oscillations at 
nearly its own eigenfrequency $\omega_{J\infty}$. That is, for $t>\tau_f$ 
the BEC and the QP subsystems are effectively decoupled. 

The mechanism for this freeze-out of BEC oscillations may now be 
interpreted as a combination of total energy conservation and 
a maximum entropy principle in the QP subsystem. The latter implies that
$n_{\text{tot}}(t)$ can essentially not decrease (up to small oscillations 
induced by the BEC driving). The energy $E_{QP}(t)$ of the QP subsystem 
increases continuously with the occupation numbers $n_m(t)$, but is 
limited by the maximum energy that can be provided by the BEC system, 
i.e., by the difference between the BEC energies in the initial and 
in the final state, $\Delta E_{BEC}=E_{BEC}(t=0)-E_{BEC}(t\to\infty)$ 
(see Supplementary Material). 
We find numerically that $E_{QP}(t)$ indeed approaches this maximum value at 
$t\approx\tau_f$. Hence, for $t>\tau_f$, 
$n_{\text{tot}}(t)$ and $E_{QP}(t)$ become approximately conserved in
the grand canonical sense, i.e., particle and energy exchange 
with the BEC are allowed, but the time averages are approximately 
constant, c.f. Fig.~\ref{qp_dynamics}. As a consequence, the 
resonant dynamics of the BEC and the QP systems must decouple, as seen 
from Fig.~\ref{spectra}. Under these dynamically generated, 
approximate conservation laws the system enters into a quasi-hydrodynamic
regime, characterized by slow, exponential relaxation, where only a 
redistribution of QPs between the individual QP levels 
occurs.

{\it Thermalization.}~--
To test if the long-time stationary state is a
thermal one, we calculate the QP distribution $b(\varepsilon_n,t)$
for different CoM times $t$. It is defined via the 
Green's functions \cite{Rammerbook} by
$F(\omega,t) = (-i/2) (2b(\omega,t)+1) A(\omega,t)$ and, hence, is 
obtained for each level from the Lorentzian weights $w_{A,n}$,
$w_{F,n}$ of these levels (c.f. Fig.~\ref{spectra}) as
\bea
b(\tilde\varepsilon_n,t)=\frac{w_{F,n}}{w_{A,n}}-\frac{1}{2}  \ .
\ea 
$\tilde\varepsilon_n$, $n=1,\, \dots,\, M$, are the 
level energies, renormalized by interactions. 
As shown in Fig.~\ref{distribution},
$b(\tilde\varepsilon_n,t)$ continuously approaches a thermal distribution.
For large $t$ this happens only by a re-distribution of weights among 
the levels, without total particle number increase. 
For even longer times the agreement with a thermal distrubution will 
be even better, since ,e.g., the occupations $n_2(t)$, $n_3(t)$ 
are still growing, while $n_5(t)$ is still decreasing 
even for the longest time shown, as seen in Fig.~\ref{qp_dynamics}.
As expected, the final-state temperature $T$ 
is high, since it is controlled by the initial BEC excitation energy,
$\Delta E_{BEC}\sim z(0)^2N_{tot}J$, which is a macroscopically large quantity.

To conclude, the system of coupled, oscillating BECs and incoherent excitations 
thermalizes, because the condensates serve as a heat reservoir for the 
quasiparticle subsystem. The condensate oscillations, in turn, get damped 
by quasiparticle collisions. By studying the system dynamics we found
a steep coupling-decoupling crossover of the condensate and the quasiparticle
subsystems at the freeze-out time scale $\tau_f$. 
Prior to $\tau_f$, the condensate and the quasiparticles are strongly coupled 
as a result of a dynamically generated parametric resonance.
For times $t>\tau_f$, BEC and incoherent excitations exhibit off-resonant 
behaviour and are effectively decoupled. This freeze-out occurs as a 
consequence of total energy conservation and entropy maximization in the 
quasiparticle subsystem.  
In the off-resonant regime, the quasiparticle system
relaxes slowly to a thermalized state with thermalization time 
$\tau_{th}\gg\tau_f$. The BEC freeze-out and subsequent time evolution under a
conservation law are reminiscent of pre-thermalization found in
low-dimensional, nearly integrable systems. However, here the
conservation law is generated dynamically in a non-integrable system.
The quasiparticle dynamics bears similarities to models for the
resonant creation and subsequent freeze-out of elementary particles  
during the evolution of the early universe.


\end{document}